\documentclass[10pt,a4paper]{article}

\pdfoutput=1

\usepackage{setspace}
\usepackage{amsmath, latexsym, amssymb}
\numberwithin{equation}{section}
\usepackage[english]{babel}
\usepackage{color}
\definecolor{myred}{rgb}{0.9,0,0}
\definecolor{myblue}{rgb}{0,0,0.8}
\definecolor{mygreen}{rgb}{0,0.8,0}
\definecolor{gray}{rgb}{0.9,0.9,0.9}
\usepackage{ifpdf}

\ifpdf
   \usepackage{graphicx}
    \usepackage[pdftex,
               pdftitle={KISS approach to credit portfolio modeling},
               pdfsubject={Risk measurement and management},
               pdfkeywords={analytical VaR, credit portfolio, capital
                 allocation, multi-factor model, KISS Model},
               pdfauthor={
               Mikhail Voropaev
               },
               pdfstartview=FitH,
               pdftoolbar=false,
               pdfmenubar=false,
               pdfwindowui=false,
               bookmarks=false,
               breaklinks=true,
               colorlinks=true,
               citecolor=mygreen,
               linkcolor=myred,
               urlcolor=myblue
               ]{hyperref}
\else
   \usepackage{hyperref}
   \usepackage[dvips]{graphicx}
   \usepackage{rotating}
\fi

\usepackage[authoryear, round, sort]{natbib}

\begin{document}
\title{KISS approach to credit portfolio modeling}
\author{Mikhail Voropaev
\thanks{ING Bank. E-mail:
\href{mailto:mikhail.voropaev@ingbank.com}{mikhail.voropaev@ingbank.com}.
}
}
\date{July 2011}
\maketitle

\begin{abstract}
A simple, yet reasonably accurate, analytical technique is proposed for multi-factor structural credit portfolio models. The accuracy of the technique is demonstrated by benchmarking against Monte Carlo simulations. The approach presented here may be of high interest to practitioners looking for transparent, intuitive, easy to implement and high performance credit portfolio model.
\end{abstract}

\section{Introduction}
Structural multi-factor \emph{economic capital} (EC) models derived
from the CreditMetrics framework \citep[][]{CreditMetrics}
have become the most widely adopted tools for risk
quantification in credit portfolios. An outcome of these models,
a portfolio EC and its allocation down to individual facilities, is used by
financial institutions for any or all of the following:  internal
capital adequacy assessment, external reporting, risk-based pricing,
performance management, acquisition/divestiture analyses,
stress-testing and scenario analysis, etc. While in most cases Monte Carlo simulations are used due to limited
analytical tractability of the multi-factor models, the recently
reported advanced analytical techniques \citep[][]{afcp} may be viewed as an alternative.

Unfortunately, neither the industry standard
simulation-based approach nor the existing analytical techniques can fully
address the needs of the financial institutions. In particular, the risk-based real-time pricing remains the ultimate challenge: none of the existing models is capable of providing sufficiently accurate, stable and time-efficient input. Yet another practical aspect which has not received enough attention in the literature is the sometimes overly complex structure of the models as perceived by end users. Very often the complexity of the models makes them hard to be understood and, hence, affects their acceptance within an organization.

The approach presented here aims to overcome the above mentioned
difficulties and is in its spirit similar to the one reported by
\citet{Cespedes}. The content, however, is quite different since the
presented model has more solid theoretical background, is easier to
implement and use and is capable of covering fully featured
multi-factor setup. The model described here was developed with KISS
principle \footnote{Keep It Simple and Straightforward,
  \href{http://en.wikipedia.org/wiki/KISS_principle}{http://en.wikipedia.org/wiki/KISS\_principle}}
in mind. While based on the previous author's research on the analytical
tractability of multi-factor models \citep[][]{afcp}, the proposed
model has very simple and intuitive structure. Despite the simple
structure, the model produces meaningful and reasonably accurate results and can be used by financial institutions for any of the purposes described above. In particular, the problem of capital allocation has a simple and time-efficient solution allowing real-time risk-based pricing. From conceptual point of view, one of the most attractive features of the model is its ability to quantify risk concentrations on both sector and obligor levels in a similar fashion.

This article is organized as follows. A short description of structural
multi-factor model and the necessary theoretical background are given
in Section \ref{sec:background}. Mathematical foundations of the proposed model are presented in
Section \ref{sec:kiss} and are substantiated by benchmarking with
Monte Carlo simulations in Section \ref{sec:benchmarking}. Section
\ref{sec:summary} contains some concluding remarks and summarizes the presentation.

\section{Background} \label{sec:background}
Let us consider a portfolio of credit risky facilities with loss
functions $\{l_i\}(\epsilon_i)$ at horizon (one year) being a function
of random variables (normalized asset returns) $\{\epsilon_i\}$.
Dependencies within the portfolio are modeled by means of a set of common
factors $\{\eta_k\}$:
\begin{eqnarray} \label{eq:Merton}
\epsilon_i = \rho_i\sum_k(\beta_i)_k\eta_k + \sqrt{1-\rho_i^2}\xi_i\cal{\beta}
\end{eqnarray}
The random variables $\{\{\eta_k\},\{\xi_i\}\}$ are independent and
standard normally distributed \footnote{In practice the common factors
are not independent are correspond to industry and geographic
sectors. However, their correlation matrix can always be
diagonalized. The latter is used here to simplify the notations.}. The
instrument specific $|\rho_i|<1$ and $\{\beta_{ik}\}$ define the systematic
sensitivities of the instruments. The latter are subject to
normalization condition $\sum_k\beta_{ik}^2=1$. The idiosyncratic risk
components are represented by $\{\xi_i\}$.

The economic capital of the portfolio is defined as na $\alpha$ - quantile
(usually set to 99.9\% or higher) of the portfolio loss distribution
$L=\sum_il_i$ relative to the expected loss of the portfolio:
\begin{eqnarray}\label{eq:ec}
\text{EC}=q_\alpha [L] - \text{E}[L]
\end{eqnarray}
The above quantifies the overall portfolio risk which can be
consistently distributed between the underlying facilities using the Euler
principle as \citep[][see e.g.]{Euler}:
\begin{eqnarray}\label{eq:euler}
\text{EC} = \sum_i\text{ec}_i, \qquad \text{ec}_i=w_i\frac{\partial}{\partial w_i}\text{EC}
\end{eqnarray}
where $w_i$ is a weight of the $i$th asset in the
portfolio. To simplify the notations, these
weights will not be written explicitly in what follows.

No closed form analytical solution exists for either portfolio EC or
its allocation $\{\text{ec}_i\}$ in general case. However, in a single-factor case,
i.e. one common factor $\eta_{1f}$ and $\rho_i=1$ for any $i$, the
portfolio loss distribution $L_{1f}$ quantile can be trivially found to
be\footnote{Here it is assumed that $L_{1f}(\eta_{1f})$ is an invertible
  function.} $q_\alpha [L_{1f}]=L_{1f}(\eta_{1f}=N^{-1}(\alpha))$, where $N^{-1}()$
is an inverse cumulative standard distribution function.
This provides motivation to look for a solution of \eqref{eq:ec} in a
form of a sum of a single factor approximation and some corrections
(for detailed explanation of this approach see e.g. \citet{afcp} and
references therein):
\begin{eqnarray}\label{eq:approx}
q_\alpha [L] = q_\alpha [L_{1f}] + \delta q_{\alpha}[\delta L_{mf}],
\qquad L_{1f} = \text{E}[L|\eta_{1f}], \quad \text{E}[L_{mf}] = 0
\end{eqnarray}
Here $L_{1f}$ is conditional on the single factor value loss
distribution of the portfolio. The single factor is constructed as a
linear combination of the systematic factors
$\eta_{1f}=\sum_k\alpha_k\eta_k$ with the normalization condition
$\sum_k\alpha^2_k=1$. Obviously, the choice of $\eta_{1f}$
significantly affects the quality of the approximation
\eqref{eq:approx} and will be given a particular attention in what
follows.

The conditional expectation series expansion technique
\citep[][]{afcp} can be applied to facilitate calculations of
$L_{1f}(\eta_{1f})$. Translated to the notations introduced here this
technique allows writing the conditional portfolio loss distribution
$L_{1f}$ as a sum of the
conditional expectations of the constituents
$\overline{l}_i=\text{E}[l_i|\eta_{1f}]$ which can be expressed as:
\begin{eqnarray} \label{eq:condexp}
\overline{l}_i(\eta_{1f}) = \sum_{n=0}^\infty
\frac{(\rho_i\vec{\beta_i}\vec{\alpha})^n}{n!}
l_i^{(n)}\text{He}_n(\eta_{1f}), \qquad l_i^{(n)}=
\int l_i(\epsilon)\text{He}_n(\epsilon)\frac{e^{-\epsilon^2/2}}{\sqrt{2\pi}}\text{d}\epsilon
\end{eqnarray}
where the inner vector product $\vec{\beta_i}\vec{\alpha}$ stands for
$\sum_n\alpha_n\beta_{in}$ and $\text{He}_n()$ are Hermite
polynomials. The series converge very well provided values of $\rho_i
\vec{\beta_i}\vec{\alpha}$ are not too close to 1 (which is the case
in practice) and allow for very fast (re)calculations of the
conditional expectations once the constants $l_i^{(n)}$ have been computed.
The above technique is particularly useful when considering arbitrary
loss functions $\{l_i\}$.

\section{KISS model} \label{sec:kiss}
\subsection{Systematic and idiosyncratic risk: happy marriage}
As long as credit portfolio modeling is concerned, it became a common
practice to distinguish the unsystematic $\{\eta_k\}$ and the
idiosyncratic $\{\xi_i\}$ risk components. The usual assumption is
that the former drive the portfolio risk dynamics while the latter
only give minor contributions. However, the two sets of random
variables do not differ from mathematical point of
view. In fact, one cannot draw a clear line between the systematic and
idiosyncratic components using practical considerations either.
Indeed, imagine that the portfolio contains a single relatively
big exposure or a set of exposures corresponding to the same borrower
and, hence, sharing the same idiosyncratic random variable
$\xi_i$. Depending on the size of this exposure(s), the risk brought
to the portfolio by $\xi_i$ may be higher than the one originating
from some or even all of the systematic factors
$\{\eta_k\}$. Big enough exposure will eventually dominate the portfolio dynamics
even if its sensitivity $\rho_i$ to the systematic factors is zero.
Introducing credit contagion effects by assigning more
than one overlapping idiosyncratic factors to a group of dependent
borrowers makes it even harder to make a distinction between the
systematic and idiosyncratic factors.

Treating the systematic and the idiosyncratic risk equally not only
simplifies the model structure, but also allows straightforward
incorporation of the borrower concentration effects into the portfolio
risk metrics. The notations used so far can be generalized as
follows. For $M$ common factors $\{\eta_k\}$ and the portfolio consisting of $N$ borrowers let us introduce
\begin{eqnarray}
\vec{\rho_i} & = & (\rho_i\beta_{i1}, \rho_i\beta_{i2}, \ldots,
\rho_i\beta_{iM}, 0, \ldots, \sqrt{1-\rho_i^2},
\ldots, 0),  \qquad \|\vec{\rho_i}\|=1 \\
\vec{\alpha} & = & (\alpha_1, \ldots, \alpha_M,
\alpha_{M+1},\ldots,\alpha_{M+N}),  \qquad \|\vec{\alpha}\|=1 \\
r_i & = & \vec{\rho_i}\cdot\vec{\alpha} = \sum_{k=1}^{M+N}\rho_{ik}\alpha_k
\end{eqnarray}
The single factor approximation $L_{1f}$ of the portfolio loss can be
written as
\begin{eqnarray} \label{eq:expansion}
L_{1f} = \sum_i\overline{l}_i, \quad
\overline{l}_i(r_i, \eta_{1f}) =
\sum_{n=0}^\infty \frac{r_i^n}{n!}l_i^{(n)}\text{He}_n(\eta_{1f})
\end{eqnarray}

Unification of the systematic and idiosyncratic risk factors results
in idiosyncratic factors being incorporated in $\vec{\alpha}$ which,
as will be shown, defines the portfolio risk dynamics. Thus, the
idiosyncratic risk is accounted for in the same fashion as the
systematic one.

\subsection{Best single factor approximation}
As was mentioned before, analytical tractability of the single factor
case is the starting point for approaching the more general
multi-factor setup. Starting with a single factor approximation and
calculating multi-factor (including idiosyncratic) adjustments as in
\eqref{eq:approx}, one can in principle calculate the portfolio
economic capital. This approach, however, suffers from some
difficulties.
First, the choice of $\eta_{1f}$ is not obvious, yet a very important
first step. Next, calculations of
the multi-factor corrections may be quite laborious and hardware
demanding. Finally, as will
be demonstrated later, the multi-factor corrections are not guaranteed
to be convergent.

Instead of trying to overcome the difficulties associated with
the multi-factor adjustments calculations, let us put all the
efforts into constructing the single-factor
approximation. Some factor $1f$ should exist which maximizes the
relative contribution of the single-factor approximation in
\eqref{eq:approx} and, thus, diminishing the relative importance of
the multi-factor corrections. Assuming that the multi-factor
corrections give positive contribution\footnote{This is not a solid
  assumption from mathematical point of view; however, it is true for
  any portfolio one may encounter in practice. In the worst case
  scenario, i.e. negative multi-factor corrections being neglected and
  the maximum single-factor contribution being used,
one is only risking being somewhat conservative.} to the $\alpha$-quantile of the
portfolio loss distribution $L$, the optimization problem reduces to
maximization of the contribution from the single-factor approximation $L_{1f}$:
\begin{eqnarray}\label{eq:maximum}
q_\alpha [L] \approx \max_{1f} q_\alpha [L_{1f}]
\end{eqnarray}
The validity of this crucial assumption will be substantiated later in
Section \ref{sec:benchmarking} by
benchmarking with Monte Carlo simulations.

From now on let us define the economic capital EC of the credit
portfolio as an $\alpha$-quantile of the optimal single factor
distribution $L_{1f}$. Using the notations introduced in this section
the economic capital can be written as
\begin{eqnarray}
\text{EC} = \sum_i\overline{l}_i(r_i)\Big|_{\eta_{1f}=N^{-1}(\alpha)}
- \sum_i\text{E}[l_i], \qquad r_i=\vec{\alpha}\cdot\vec{\rho_i}
\end{eqnarray}
The optimal single factor is defined by $\vec{\alpha}$ which maximizes
the above expression
\begin{eqnarray}
\nabla_{\vec{\alpha}}\sum_i\overline{l}_i(r_i) = 0
\end{eqnarray}
and has the following solution
\begin{eqnarray}\label{eq:optimal}
\vec{\alpha} = \frac{\vec{p}}{\|\vec{p}\|}, \qquad \vec{p} = \sum_i\frac{\partial\overline{l}_i(r_i)}{\partial
  r_i}\vec{\rho}_i
\end{eqnarray}
This equation, however, contains $\vec{\alpha}$ on both sides ($r_i$ on the
right contains $\alpha$) and does not allow a straightforward analytical
solution. The problem \eqref{eq:maximum} can still be solved numerically by applying, for
example, the method of steepest descent. Based on \eqref{eq:optimal},
the following starting point can be suggested
\begin{eqnarray}\label{eq:start}
\vec{\alpha}_0 = \frac{\vec{p}_0}{\|\vec{p}_0\|}, \qquad \vec{p}_0 =
\sum_i\frac{\partial\overline{l}_i(r_i)}{\partial r_i}\vec{\rho}_i\Big|_{r_i=0}
\end{eqnarray}
The calculations can be significantly facilitated by the series
expansion \eqref{eq:expansion}. In practice, only few iterations are
needed to have an accurate solution to the optimization problem
\eqref{eq:maximum}. The calculations are not hardware demanding and
very fast.

The optimal single factor defined by \eqref{eq:optimal} leads to
another simplification for the portfolio capital allocation problem
\eqref{eq:euler}. The individual capital contributions
\begin{eqnarray}
\text{ec}_i = w_i\frac{\partial}{\partial w_i}\text{EC} = w_i\frac{\partial}{\partial
  w_i}\sum_i\left(\overline{l}_i(r_i) - \text{E}[l_i]\right)
\end{eqnarray}
can be written as
\begin{eqnarray}\label{eq:capalloc}
\text{ec}_i = \overline{l}_i - \text{E}[l_i] +
\sum_j\frac{\partial\overline{l}_i}{\partial r_i}\vec{\rho}_j\cdot
w_i\frac{\partial}{\partial w_i}\frac{\vec{p}}{\|\vec{p}\|} =
\overline{l}_i(\vec{\alpha}\vec{\rho}_i) - \text{E}[l_i]
\end{eqnarray}
where the third term can be shown to be zero:
\begin{eqnarray}
\sum_j\frac{\partial\overline{l}_j}{\partial r_j}\vec{\rho}_j\cdot
w_i\frac{\partial}{\partial w_i}\frac{\vec{p}}{\|\vec{p}\|} =
\vec{p}\cdot w_i\frac{\partial}{\partial
  w_i}\frac{\vec{p}}{\|\vec{p}\|} =
\vec{p}\cdot\left(\frac{\vec{\rho}_i}{\|\vec{p}\|}-
\frac{\vec{p}(\vec{p}\cdot\vec{\rho}_i)}{\|\vec{p}|^3}\right) = 0
\end{eqnarray}
In other words, the choice of $\vec{\alpha}$ according to
\eqref{eq:optimal} leads to particularly simple expressions for
capital contributions \eqref{eq:capalloc}. The overall portfolio EC is
a sum of conditional expectations of the excess losses
$\overline{l}_i(\eta_{1f}=N^{-1}(\alpha)) - \text{E}[l_i]$ which happen to
coincide with the individual capital contributions.

Let us emphasize that once the optimal single factor
\eqref{eq:optimal} has been computed, the set of parameters
$\vec{\alpha}$ is sufficient to perform the capital allocation
calculations \eqref{eq:capalloc}. This allows real time calculations
necessary for risk-based pricing. In fact, the proposed method for the
economic capital calculations and its allocation is so efficient that
the calculations can be performed on huge portfolios containing
millions of facilities in minutes using entry level desktop
computer. The accuracy of the method is demonstrated in the next
section through benchmarking against Monte Carlo simulations.

\section{Benchmarking} \label{sec:benchmarking}
To highlight the points made and substantiate the assumptions used in
the previous section, let us compare the performance (accuracy) of the
proposed analytical approximation with Monte Carlo simulations. To
demonstrate the advantages of the proposed technique, the comparison
analysis will also cover the previously reported
analytical technique \citep[][]{afcp} which, in contrast with the one
presented here, aims at precise calculations of the multi-factor
corrections in \eqref{eq:approx}.

\begin{figure}[h!]
\centering
\ifpdf
$\begin{array}{l}
\includegraphics[width=0.4\textwidth,viewport=0 600 265 790,clip]{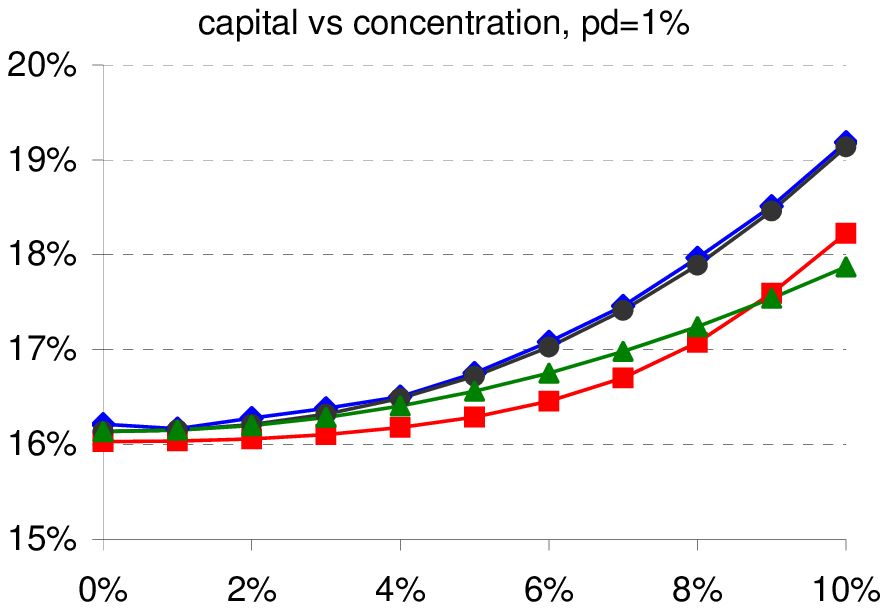}
\includegraphics[width=0.4\textwidth,viewport=0 595 265 785,clip]{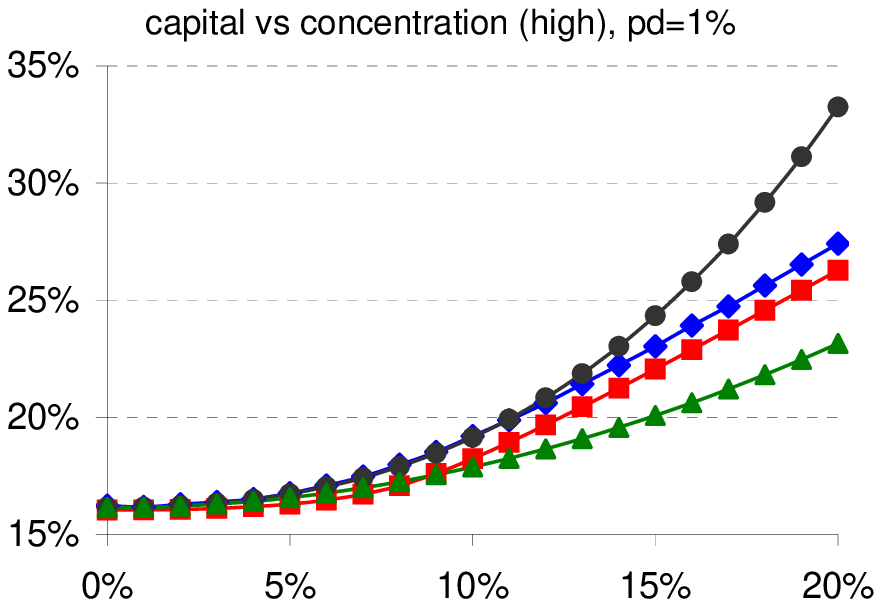} \\
\includegraphics[width=0.4\textwidth,viewport=0 600 265 790,clip]{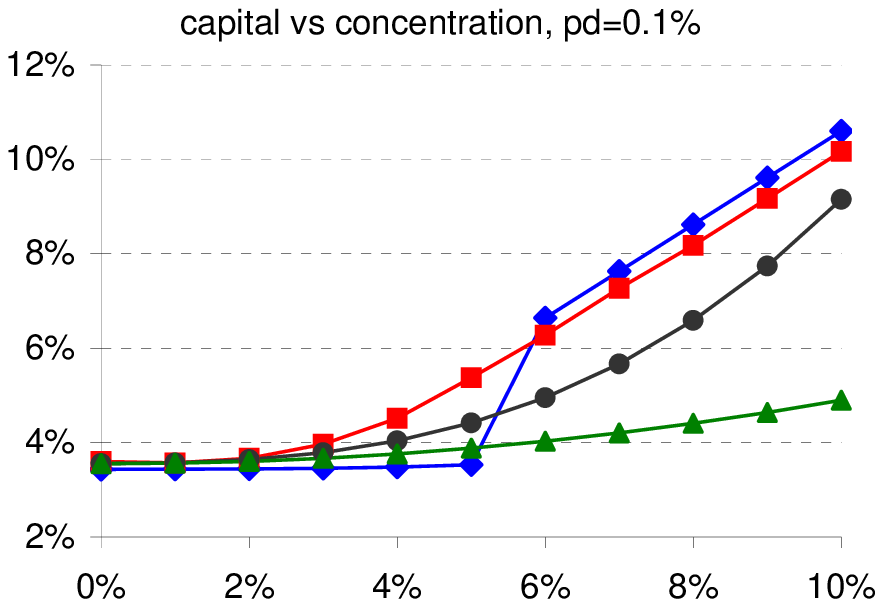}
\includegraphics[width=0.4\textwidth,viewport=0 595 265 785,clip]{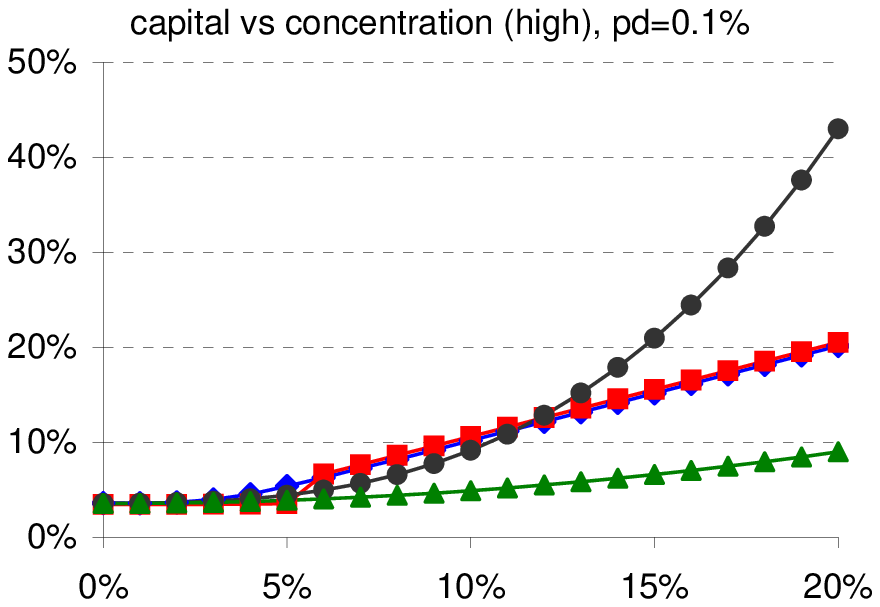} \\
\includegraphics[width=0.8\textwidth,viewport=0 600 530 790,clip]{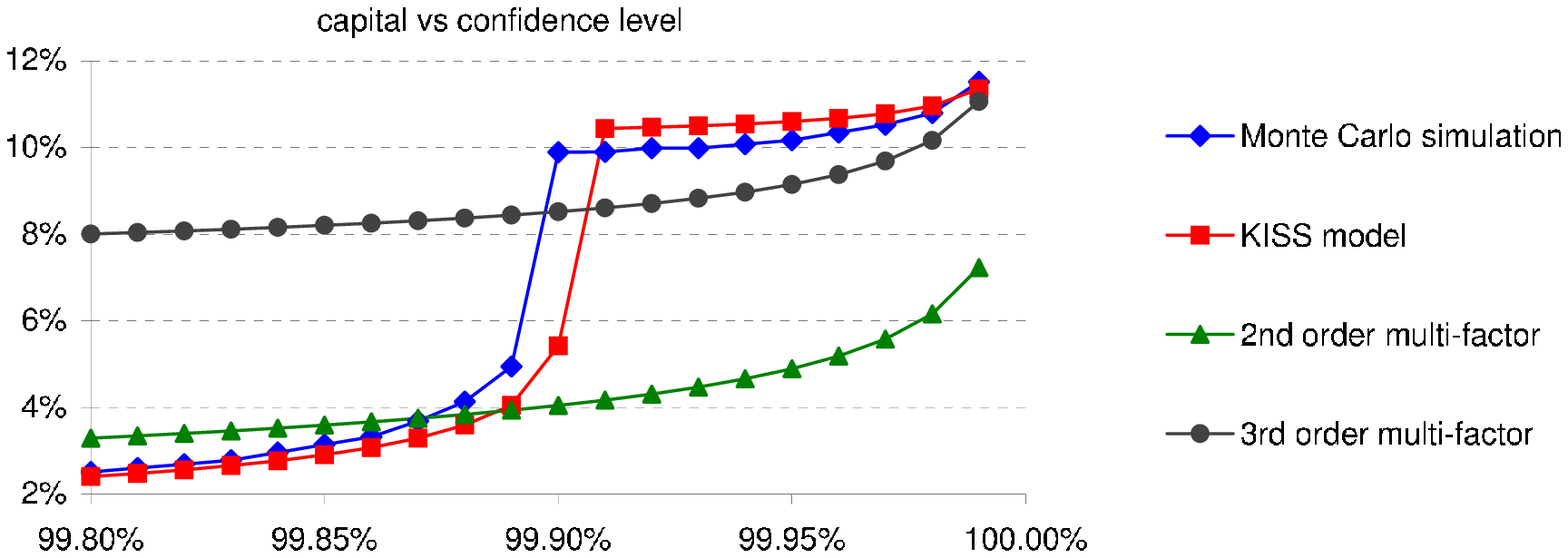}
\end{array}$
\else
$\begin{array}{ccc}
\includegraphics[width=0.4\textwidth,viewport=0 600 265 790,clip]{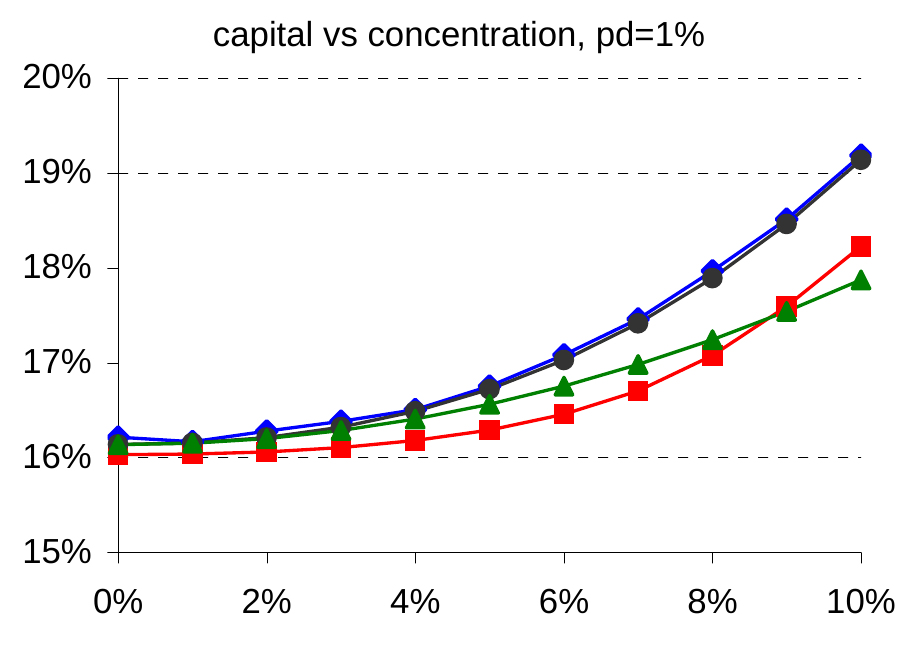}
\includegraphics[width=0.4\textwidth,viewport=0 595 265 785,clip]{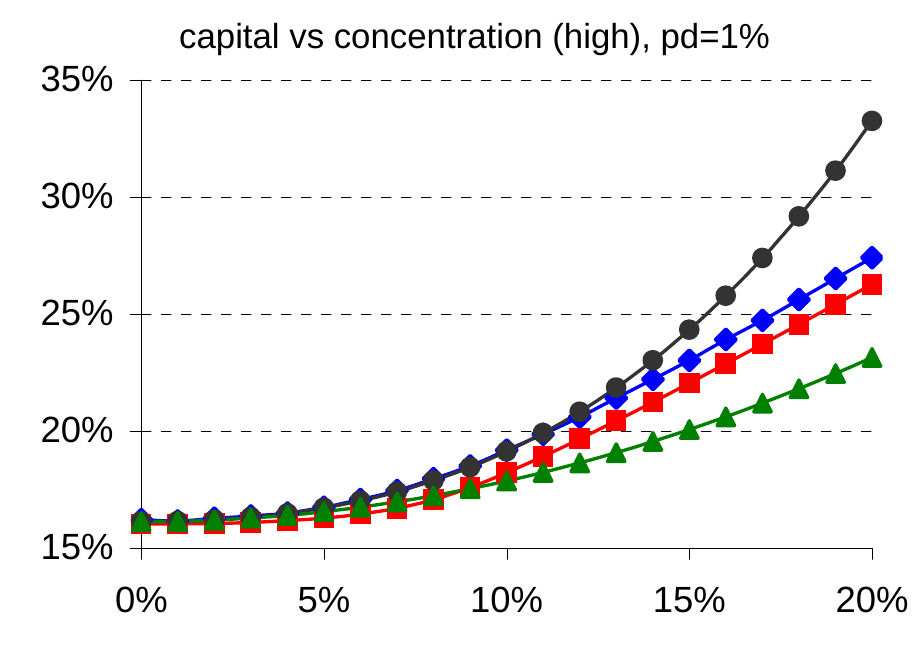} \\
\includegraphics[width=0.4\textwidth,viewport=0 600 265 790,clip]{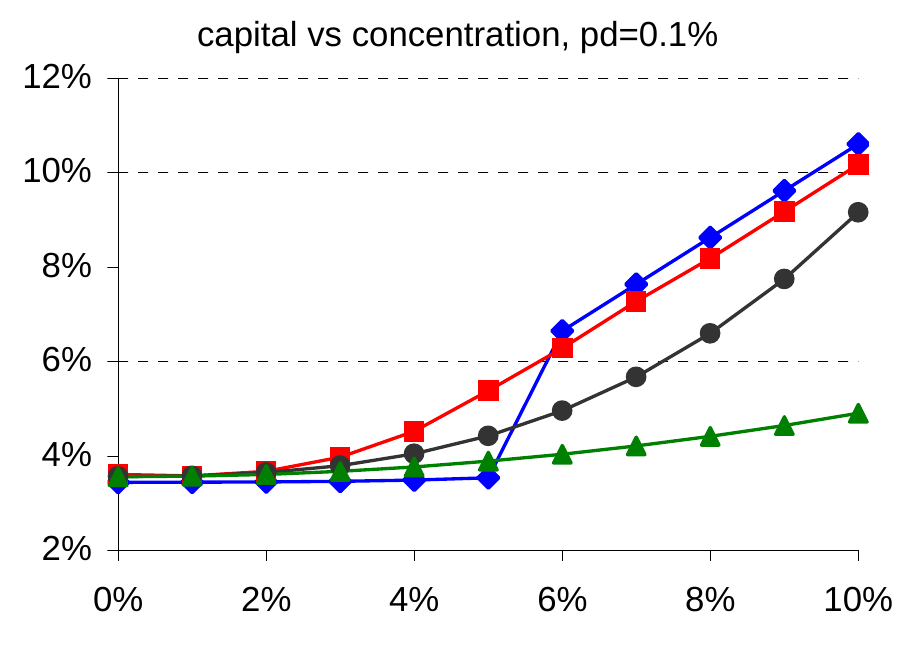}
\includegraphics[width=0.4\textwidth,viewport=0 595 265 785,clip]{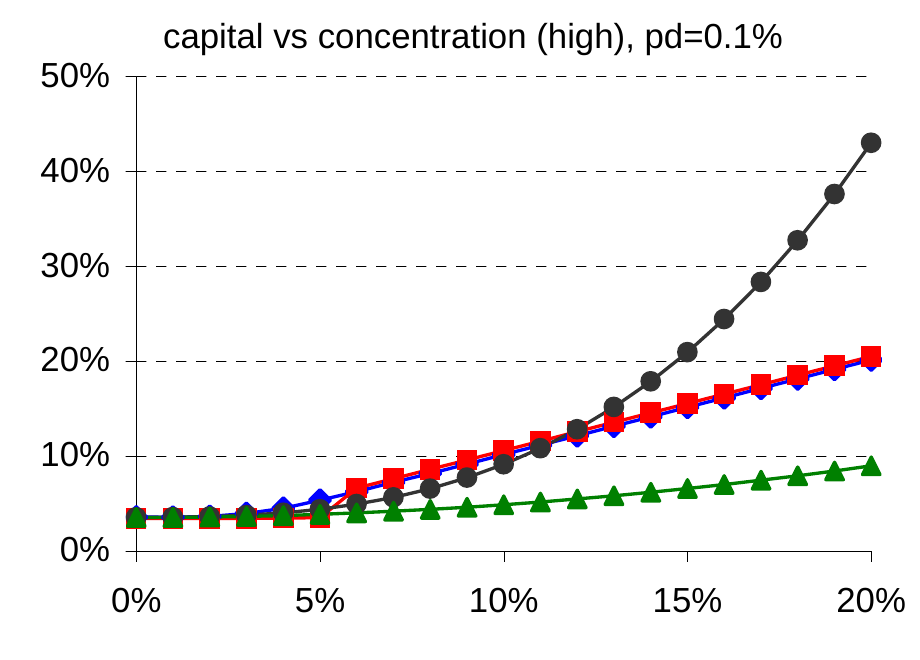} \\
\includegraphics[width=0.8\textwidth,viewport=0 0 530 190,clip]{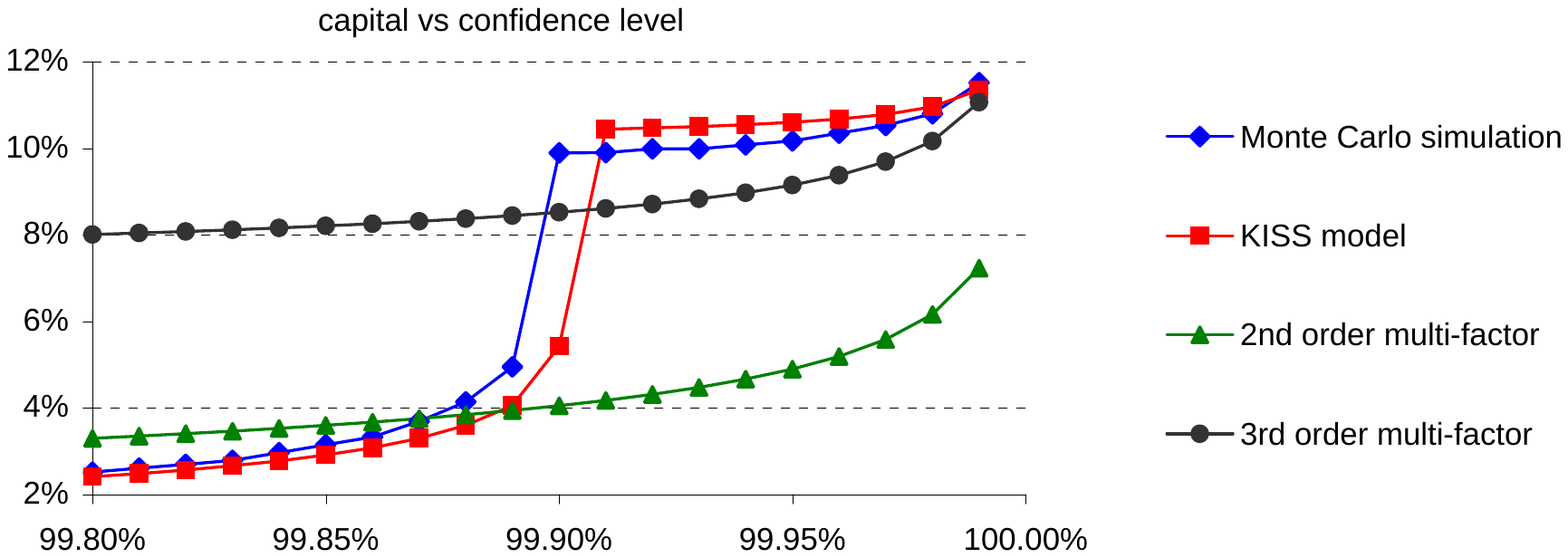}
\end{array}$
\fi
\parbox{0.75\textwidth}{
\caption{\emph{Portfolio EC: Monte Carlo vs. analytical estimates.}\label{fig:artificial}}}
\end{figure}

\subsection{Artificial portfolio, high concentrations}
Let us start the analysis by considering a simple portfolio of 1000
bullet loans maturing at the horizon. Each loan has sensitivity
$\rho_i^2=0.2$ to the single systematic factor $\eta$ and a unique
idiosyncratic component $\xi_i$. The loss at horizon
functions are
\begin{eqnarray}
l_i(\epsilon_i) = \left\{
\begin{array}{rl}
l_0 & \text{if } \epsilon > -N^{-1}(\text{PD}) \\
0 & \text{if } \epsilon \leq -N^{-1}(\text{PD})
\end{array}
\right.
\end{eqnarray}
where $\text{PD}_i$ are \emph{probabilities of default} and are equal
for all the
loans. Two sets of experiments were conducted with PD=1\% and
PD=0.1\%. The loss severities $l_0$ are initially set equal to all loans in the portfolio.
The concentration effects are studied by gradually increasing the loss severity
of one of the loans and examining the impact on the portfolio EC. The
confidence level for EC is set to 99.9\% except for the last
experiment where the dependency of the EC on the confidence level was
investigated for fixed 10\% concentration. Both EC and concentration are measured as a fraction of the total
portfolio exposure (i.e. the sum of all the loss severities).

The results of the Monte Carlo simulations are compared with
the numbers produced using the KISS model (Section \ref{sec:kiss}) and with the
output of the approach based on the calculations of the multi-factor
corrections in \eqref{eq:approx}. In the latter case the single
systematic factor $\eta$ was used for the single-factor approximation.

Based on the results of the comparison presented in
Fig. \ref{fig:artificial}, one can conclude the
following. Multi-factor corrections may lead to very accurate results
in case of moderate concentrations (PD=1\%, concentration $<$ 10\%). In
case of high concentrations, however, the results suggest that the higher
order (3rd order in this case) multi-factor corrections are divergent, while limiting the
multi-factor contributions to the 2nd order corrections leads to significant
underestimation of EC.

The KISS model, on the other hand, despite
being not extremely accurate, produces robust and reliable EC
estimates for a wide range of parameters. This is especially obvious
when studying the dependency of the portfolio EC on the confidence
level.

\subsection{Realistic portfolio, moderate concentrations}
Here the analysis of the previous section is complemented by the one
conducted on more realistic portfolio.
The portfolio consisted of 2,000 loans to distinct customers randomly
selected from a loan portfolio of a large European bank\footnote{The
  same portfolio was used by \citet{afcp}.}. The set of common systematic factors covering 45 geographic regions and 61 regions, as well as the valuation function at horizon $l_i(\epsilon)$ used in the experiment were similar to those of the PortfolioManager \citep[][]{PortfolioManager} model.

Both the portfolio EC and its allocation $\{\text{ec}_i\}$ were
estimated using unbiased Monte Carlo simulations. The confidence level
was set to 99.9\% and the EC contributions were estimated as average
realized values in the interval 99.85\%-99.95\%. A total of $10^{10}$
(ten billion) scenarios were used.

The simulation-based estimates were compared to both the KISS model
outcome as well as with the
output of the approach based on the calculations of the multi-factor
corrections in \eqref{eq:approx}.
In the latter case the single-factor $1f$ used as a
starting point was obtained using \eqref{eq:start} restricted to
systematic components. Second ($mf2$) and third order ($mf3$) corrections were
calculated.

\begin{figure}[t!]
\centering
\ifpdf
$\begin{array}{ccc}
\includegraphics[width=0.45\textwidth,viewport=10 585 390 785,clip]{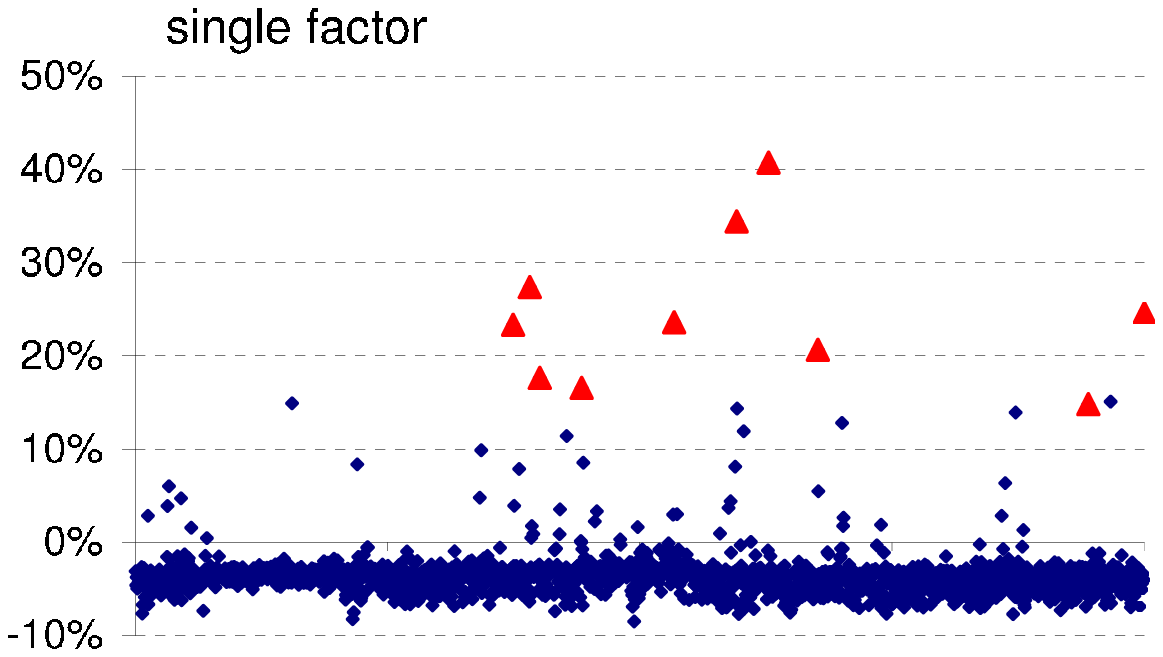} &
\includegraphics[width=0.45\textwidth,viewport=5 590 385 790,clip]{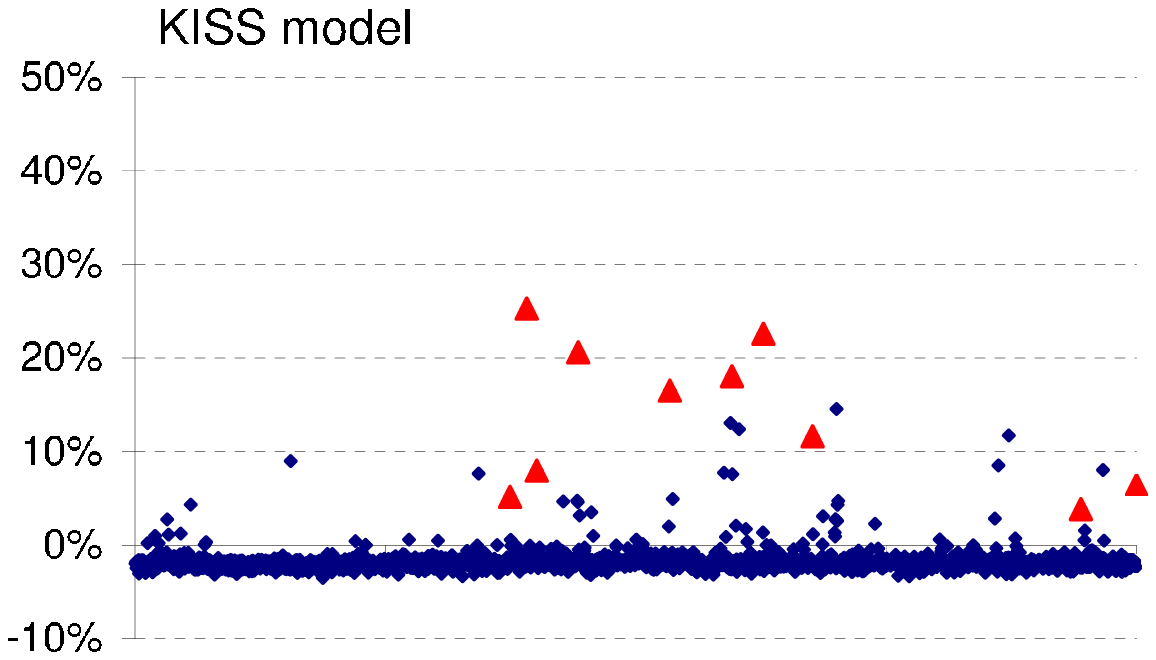} \\
\includegraphics[width=0.45\textwidth,viewport=5 590 385 790,clip]{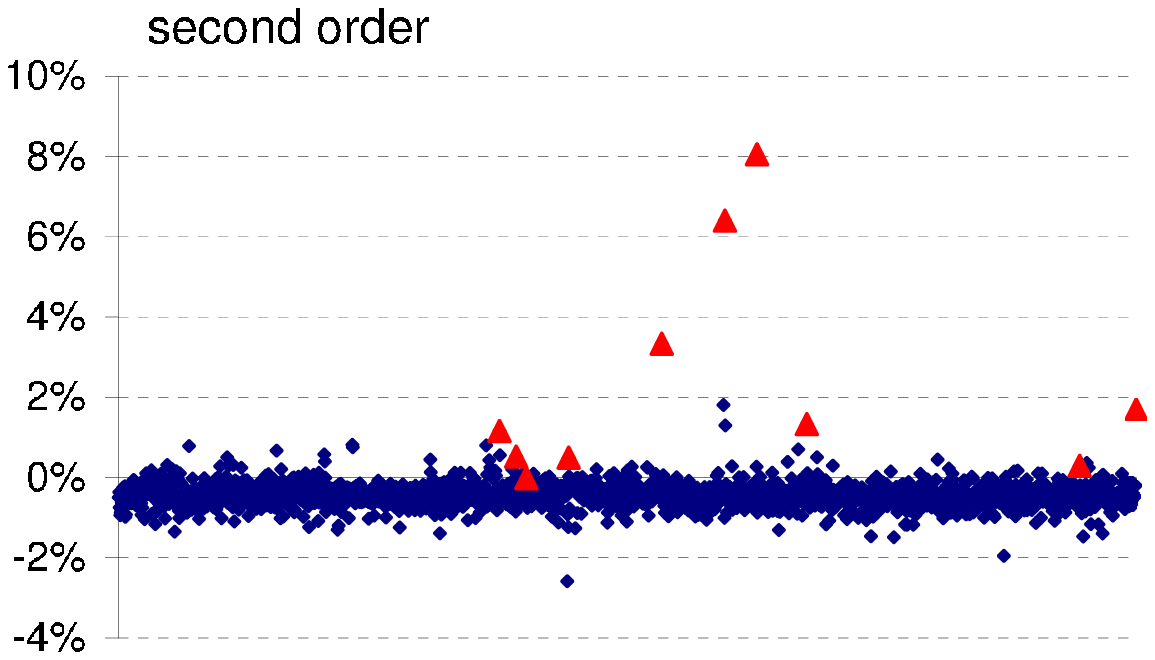} &
\includegraphics[width=0.45\textwidth,viewport=5 590 385 790,clip]{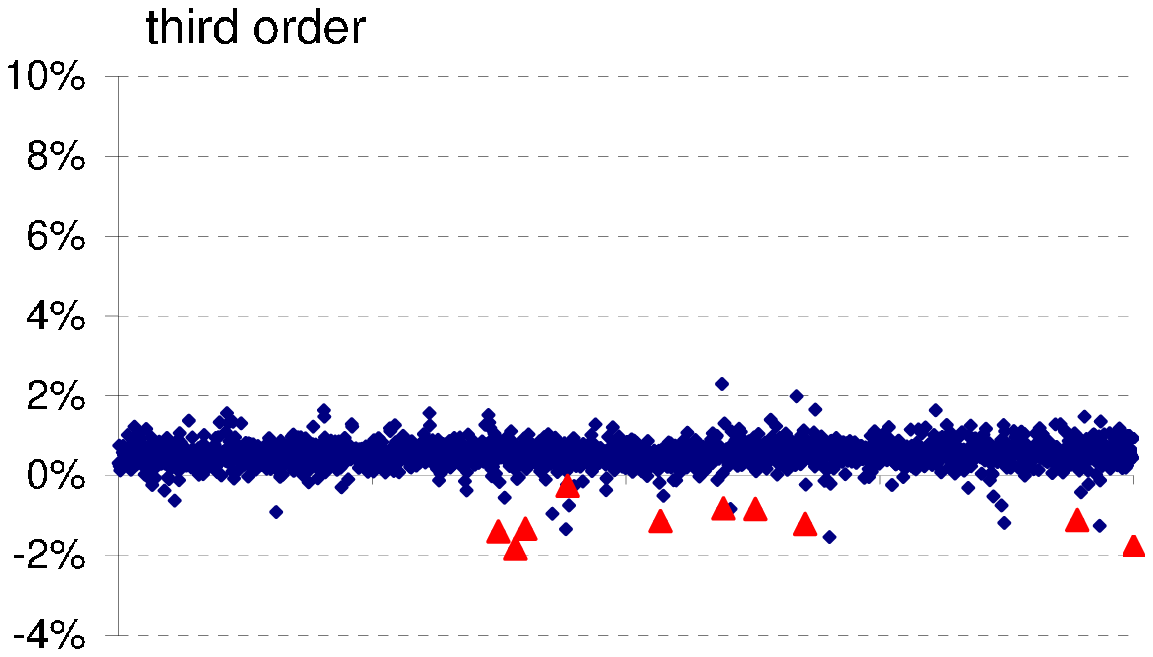}
\end{array}$
\else
$\begin{array}{ccc}
\includegraphics[width=0.45\textwidth,viewport=0 5 380 205,clip]{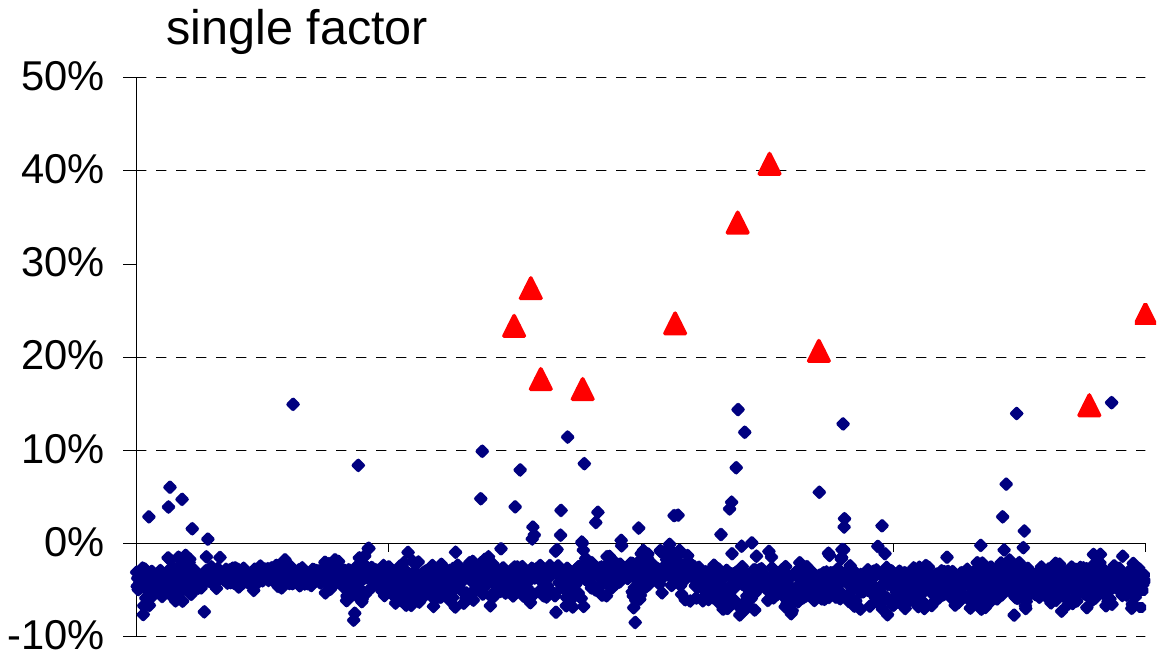} &
\includegraphics[width=0.45\textwidth,viewport=0 5 380 205,clip]{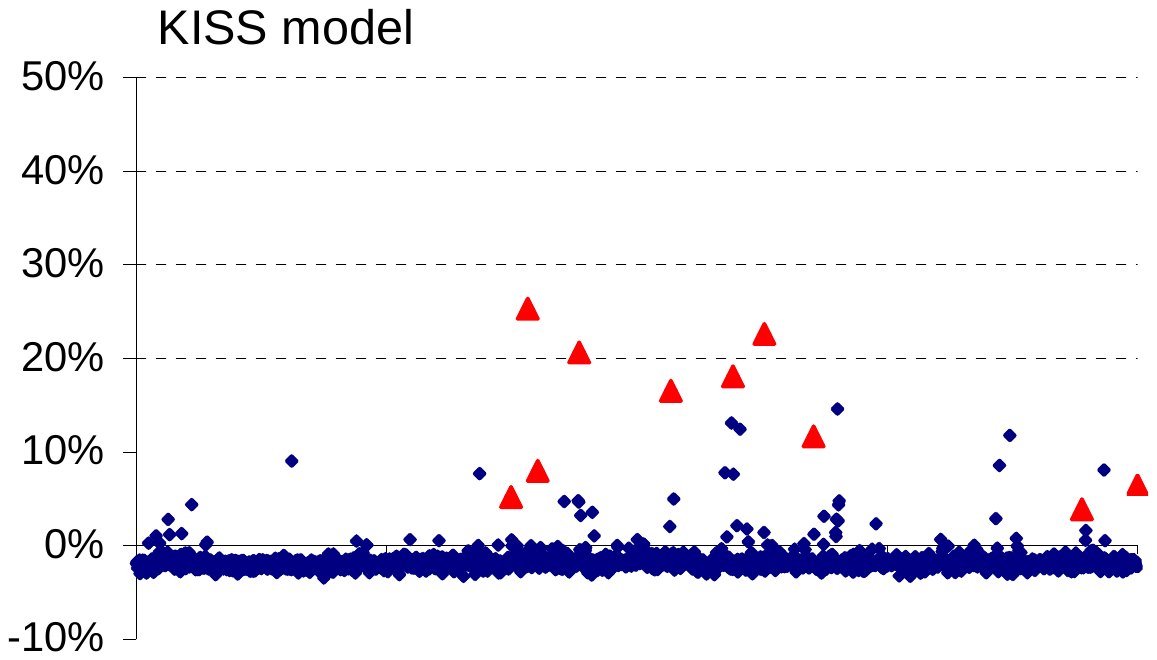} \\
\includegraphics[width=0.45\textwidth,viewport=0 5 380 205,clip]{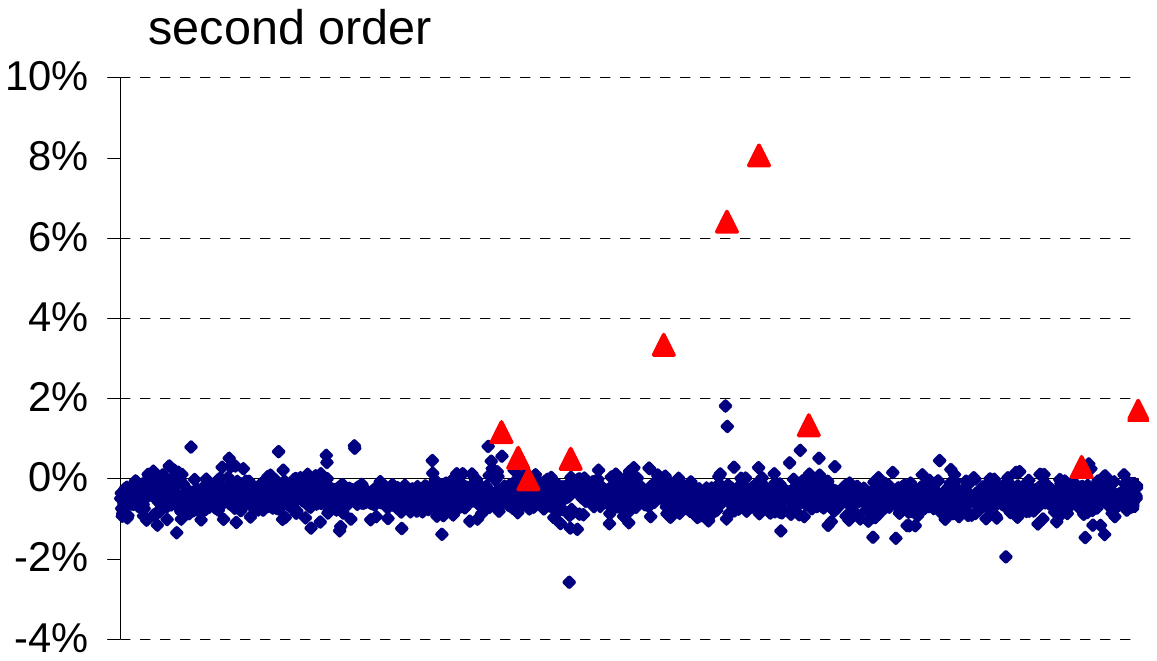} &
\includegraphics[width=0.45\textwidth,viewport=0 5 380 205,clip]{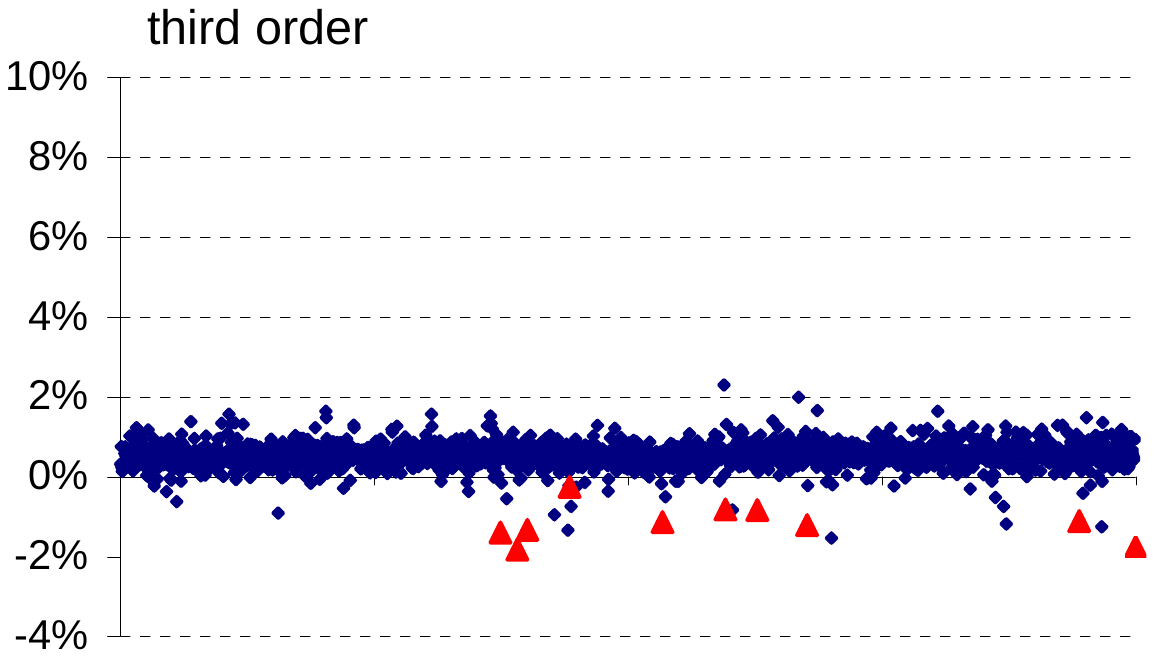}
\end{array}$
\fi
\parbox{0.75\textwidth}{
\caption{\emph{Relative differences between Monte Carlo and analytical estimates. Ten biggest consumers of capital accounting for 19\% of EC are marked.}\label{fig:realistic}}}
\end{figure}

The results on the portfolio level are summarized in Table
\ref{tab:portfolio} and the facility-level results are presented as
scatter plots in Fig.\ref{fig:realistic}.

\begin{table}[h!]
\centering
\begin{tabular}{c c c c}
\hline\hline
1f & KISS & 1f+mf2 & 1f+mf2+mf3  \\ \hline \\
-4.3\% & -2.0\% & -0.5\% & -0.1\% \\ \hline
\end{tabular}
\parbox{0.6\textwidth}{
\caption{\emph{Relative differences between analytical and simulation-based estimates of the portfolio EC.}}
\label{tab:portfolio}}
\end{table}

On the portfolio level the accuracy of the proposed approximation is
more than sufficient for any practical purposes. The performance may
seem to be less impressive on the facility level. As expected, the
most significant mismatch is observed for the facilities with high
concentrations of capital. These discrepancies, however, do not
jeopardize the practical validity of the KISS model. Indeed, the
uncertainty in the input parameters observed in practice (i.e. default
rates, recoveries, correlation parameters, etc) as well as
dependencies on the modeling assumptions diminish to large extent the
approximation errors of the proposed technique. This, combined with the robustness demonstrated in the previous section, makes the KISS approximation a valid alternative for quantification of risks in credit portfolios.  

\section{Summary} \label{sec:summary}
Despite its simplicity, the analytical approximation presented here is
capable of quantifying credit portfolio risks in a general
multi-factor setup. The \emph{VaR} risk measure used here can easily be replaced
with the \emph{Expected Shortfall}. The arbitrary loss functions $\{l_i(\epsilon_i)\}$
used allow for covering not only default-only regime, but also MtM
valuation or even the dependency of in-default loss severities on the
systematic factors. The default-only case has a particularly simple
solution mimicing the well-known IRB capital rules:
\begin{eqnarray}
\text{ec}_i = \text{EaD}_i\cdot\text{LGD}_i\cdot\left( 
N\!\left(\frac{N^{-1}(\text{PD}_i)+(\vec{\alpha}\vec{\rho}_i)\,
    N^{-1}(\alpha)}{\sqrt{1-(\vec{\alpha}\vec{\rho}_i)^2}}\right)-\text{PD}_i\right)
\end{eqnarray}

The less than perfect accuracy of the approximation is not crucial for day-to-day practical needs of credit portfolio managers. The advantages of the proposed technique are significant. the model allows very fast and straightforward calculations including real-time risk-based pricing. The simple, robust and transparent structure can facilitate user acceptance and integration on all levels of financial institutions.

\end{document}